\documentclass[aps,pra,amssymb, amsmath,nobibnotes,nobibnotes,  superscriptaddress, showpacs,showkeys,twocolumn,floatfix]{revtex4-1}
\usepackage{graphicx}
\usepackage{epstopdf}
\usepackage{graphicx,graphics,color,epsfig}
\usepackage{revsymb4-1}
\usepackage{bm}
\usepackage{braket}
\usepackage{mathptmx}
\usepackage[colorlinks, linkcolor=blue, anchorcolor=blue, citecolor=blue]{hyperref}
\usepackage{hyperref}

\begin{document}

\title{Quadrupole absorption rate and orbital angular momentum transfer for atoms in optical vortices} 

\author{Smail Bougouffa \thanks{corresponding author}}
\email{sbougouffa@hotmail.com; sbougouffa@imamu.edu.sa}
\affiliation{Department of Physics, College of Science, Imam Mohammad ibn Saud Islamic University (IMSIU), P.O. Box 90950, Riyadh 11623, Saudi Arabia\\
ORCiD:  http://orcid.org/0000-0003-1884-4861}

\author{ Mohamed Babiker}
\email{m.babiker@york.ac.uk }
\affiliation{Department of Physics, University of York, Heslington,York, YO10 5DD, U.K.\\
ORCiD:  http://orcid.org/0000-0003-0659-5247}

\date{\today}

\begin{abstract}
Recent experiments involving the interaction of optical vortices with atoms in quadrupole transitions have been shown to be accompanied by the exchange of orbital angular momentum (OAM) between the electronic states of the atom and the optical vortex field.  Earlier work by both theory and experiment had ruled out the transfer of a vortex OAM to the electronic degrees of freedom in an electric dipole atomic transition and it has been confirmed that the lowest multipolar order involving an OAM transfer to the electronic motion is indeed the electric quadrupole. Hitherto, the quadrupole  transition involving optical vortices has not been quantified and we have thus set out to evaluate the absorption rate accompanied by an OAM transfer with reference to the $6^2S_{1/2}\rightarrow 5^2D_{5/2}$ in Cs when caesium atoms are subject to the field of a linearly polarized optical vortex. Our results assuming typical experimentally accessible parameters indicate that the absorption rate for moderate light intensities is smaller than the quadrupole spontaneous emission  rate, but should still be within the measurement capabilities of modern spectroscopic techniques. 
\end{abstract}



\maketitle

\section{Introduction}\label{sec1}
Twisted light beams or optical vortices have been the subject of much investigation over the last three decades or so and this area has found  applications in a number of diverse fields \cite{Yao2011a,TorresTorner2011,Surzhykov2015}. The absorption or emission of any type of light by atoms is accompanied by the transfer of linear momentum between the light and the atoms, and this effect has been exploited in the cooling and trapping of atoms \cite{metcalflaser, FrankeArnold2008, babiker2018atoms}. It has also been established that the application of vortex light to atoms affects the centre of mass motion \cite{babiker2018atoms,lembessis2014atom} via the dipole force, but the quadrupole force too can influence atomic motion attracting the atoms to the high or low-intensity regions of the light field \cite{ Al-Awfi2019,Ray2020a, Bougouffa20}.

The possibility of exchanging orbital angular momentum (OAM) between light and the internal motion of atoms, as opposed to their gross, or centre of mass, motion, has interested researchers since the early 1990s \cite{van1990light, van1994selection}.
The first treatments to tackle the problem of transfer \cite{29} concluded that while the centre of mass of atoms engage with an optical vortex, the internal `electronic-type' degrees of freedom of the atom do not take part in any OAM exchange in an electric dipole transition and that only in a quadrupole interaction that  an exchange involving the electronic (internal)  degrees of freedom can take place.  This result was subsequently confirmed experimentally by a number of researchers, first by Araoka et al \cite{Araoka2005} who showed that optical vortex light is not specific in the interaction with chiral matter and then by L\"{o}ffler et al \cite{loffler2012cholesteric,Loeffler2011a} whose experimental work could not detect any influence of the OAM of circular dichroism in cholestric polymers. The recent experimental work by Giammanco et al \cite{Giammanco2017} confirmed categorically the lack of influence of the OAM on electric dipole transitions in agreement with theory \cite{29,Araoka2005}.

Research has also dealt with the study of the mechanical forces on atoms due to the coupling of optical vortices, such as Laguerre-Gaussian and Bessel-Gaussian modes, to quadrupole-active atomic transitions\cite{Bougouffa20, lembessis2013enhanced,babiker2018atoms} and the results displayed a considerable enhancement in the case of twisted beams due to the gradient coupling which increases with increasing winding number. Quadrupole transitions have also been observed in the case of Rb atoms by evanescent light when Rb is localized in the vicinity of an optical nano-fibre \cite{Ray2020a}.

As a potentially measurable effect the absorption of vortex light and the rate of absorption by atoms has not been evaluated as far as the authors are aware.  We have therefore set out to evaluate the rate of absorption which in the case of an optical vortex is also interpretable as the rate of OAM transfer from the vortex light to the atoms.  The general treatment is then applied to the particular case involving the $6^2S_{1/2}\rightarrow 5^2D_{5/2}$ quadrupole transition in Cs when caesium atoms are subject to the field of an optical vortex.  The Cs transition in question is well-known as a dipole-forbidden but quadrupole allowed transition. 

This paper is organized as follow. The basic theory is outlined in Sec.\ref{sec2}, while Sec.\ref{sec3} presents the theory of the quadrupole atomic absorption rate when the atom engages with the optical vortex field at near-resonance.  The evaluation involves a direct application of the Fermi Golden rule where the  usual selection rules are applicable for quadrupole transitions , but absorption requires a treatment involving the density of of the continuum states as a Lorentzian function representing the upper atomic level as an energy band of width $\hbar\gamma$ where $\gamma^{-1}$ is the lifetime of the upper state. Section \ref{sec4} deals with the case where the applied vortex beam is a Laguerre-Gaussian beam.
The results are illustrated in Sec.\ref{sec5} for the quadrupole atomic transition $6^2S_{1/2}\rightarrow 5^2D_{5/2}$ in Cs. A summary of our results and brief comments on their significance are given in Sec.\ref{sec6}.

\section{Quadrupole Rabi frequency }\label{sec2}

The quantum system consists of the two-level-atom interacting with a single optical vortex beam propagating along the $+z$ axis. The ground and excited states of the two-level-atom are $\{\ket{g}, \ket{e}\}$ with level energies $\mathcal{E}_1$ and $\mathcal{E}_2$, respectively, which correspond to a transition frequency $\omega_a=(\mathcal{E}_2- \mathcal{E}_1)/\hbar$. The interaction Hamiltonian is written as a multipolar series expansion about the center of mass coordinate $\mathbf{R}$ as follows \cite{Bougouffa20,babiker2018atoms,lembessis2013enhanced,Al-Awfi2019}
\begin{equation}\label{1}
    \hat{H}_{int}=\hat{H}_{dp}+\hat{H}_{qp}+...,
\end{equation}
where the first term  $\hat{H}_{dp}=- \hat{\bm{\mu}}.\mathbf{ \hat{E}}(\mathbf{R})$  stands for the electric dipole interaction between the atom and the electric field,  
$\bm{ \hat{\mu}}=e\mathbf{r}$, with $\mathbf{r}$ the internal position vector, is the electric dipole moment vector and $\bm{ \hat{E}}(\mathbf{R})$ is the electric field vector. The optical transition in question is taken here to be dipole-forbidden, but quadrupole-allowed, so it is the second (quadrupole) interaction term in Eq.(\ref{1})that is of relevance here.  We have
\begin{equation}\label{2}
 \hat{H}_{qp}=-\frac{1}{2}\sum_{ij} \hat{Q}_{ij} \nabla_i \hat{E_j}.
\end{equation}
Here $x_i$ are the components of the internal position vector $\mathbf{r}=(x, y, z)$ and $\nabla_i$ are components of the gradient operator which act only on the spatial coordinates of the transverse electric field vector $\mathbf{ E}$ as a function of the centre of mass position vector variable $\mathbf{ R}= (X, Y, Z)$.
The quadrupole tensor operator  ${\hat{Q}}_{ij}$  can be written in terms of ladder operators as $ \hat{Q}_{ij}=Q_{ij}( \hat{\pi} + \hat{\pi}^{\dag})$, where $Q_{ij}=\bra{i}\hat{Q}_{ij}\ket{j}$ are the quadrupole matrix elements between the two atomic levels, and $ \hat{\pi} ( \hat{\pi}^{\dag})$ are the atomic level lowering (raising) operators.

Without loss of generality, we assume that the electric field is plane-polarized along the $x$ direction, so optical spin has no role to play here, in which case  we have the following form of the quadrupole interaction Hamiltonian
\begin{equation}\label{3}
  \hat{H}_{qp}=-\frac{1}{2}\sum_{i} \hat{Q}_{ix}\frac{\partial  \hat{E_x}}{\partial R_i}
\end{equation}
The quantized electric field can conveniently be written in terms of the centre-of-mass position vector in cylindrical polar coordinates $\mathbf{R}=(\rho,\phi, Z)$ as follows
\begin{equation}\label{4}
    \mathbf{ \hat{E}}(\mathbf{R})=\mathbf{ \hat{i}} u_{\{k\}}(\mathbf{R})\hat{a}_{\{k\}}e^{i \theta_{\{k\}}(\mathbf{R})}+H.c.
\end{equation}
where $u_{\{k\}}(\mathbf{R})$ and $\theta_{\{k\}}(\mathbf{R})$ are, respectively, the amplitude function and the phase function of the LG vortex electric field. Here the subscript $\{k\}$ denotes a group of indices that specify the optical mode in terms of its axial wave-vector $k$, winding number $\ell$ and radial number $p$. The operators $\hat{a}_{\{k\}}$ and $\hat{a}_{\{k\}}^{\dagger}$ are the annihilation and creation operators of the field mode $\{k\}$. Finally $H.c.$ stands for Hermitian conjugate.
Using this form of the electric field, we obtain the desired expression for the quadrupole interaction Hamiltonian 
\begin{equation}\label{5}
    \hat{H}_{qp}=\hbar\Omega^{Q}_{\{k\}}(\mathbf{R})e^{i\theta_{\{k\}}(\mathbf{R})}\hat{a}_{\{k\}}(\hat{\pi}^{\dag}+\hat{\pi})+H.c.
\end{equation}
where $\Omega^{Q}_{\{k\}}(\mathbf{R})$ is the quadrupole Rabi frequency, which can be written as
\begin{equation}\label{5p}
   \Omega^{Q}_{\{k\}}(\mathbf{R})=-\frac{1}{2\hbar}\sum_{i} Q_{ix} u_{\{k\}}\Big( \frac{1}{ u_{\{k\}}}\frac{\partial u_{\{k\}}}{\partial R_i}+i\frac{\partial \theta_{\{k\}}}{\partial R_i} \Big)
\end{equation}

It is convenient to proceed as we show below by assuming a general LG mode LG$_{\ell p}$ of winding number $\ell$ and radial number $p$.  The values of $\ell$ and $p$ applicable to a given quadrupole transition is decided by application of the selection rules of the specific atomic transition.

\section{Fermi Golden rule for absorption rate}\label{sec3}
The vortex field is endowed with an orbital angular momentum $\pm \ell \hbar$ per photon with $\ell$ positive.
Thus, the transition matrix element \cite{forbes2018chiroptical}, including only the quadrupole coupling, is given by
\begin{equation}\label{6}
 \mathrm{T}^{\{k\}}_{if}=\bra{f}\hat{H}_{qp}\ket{i},
\end{equation}
where $\ket{i}$ and $\ket{f}$ are, respectively, the initial and final states of the overall quantum system (atom plus optical vortex). We assume that the system has as an initial state $\ket{i}$ with the atom in its the ground state and there is one vortex photon.  The final state $\ket{f}$  consists of the excited state of the atom and there is no field mode. Thus $\ket{i}=\ket{g\{1\}_{\{k\}}}$ and $\ket{f}=\ket{e\{0\}}$.  
\begin{equation}\label{7}
\mathrm{T}^{\{k\}}_{if}=-\frac{1}{2}\sum_{ij}\bra{e}\hat{Q}_{ij}\ket{g} \bra{\{0\}} \frac{\partial  \hat{E_j}}{\partial R_i} \ket{\{1\}_{\{k\}}} .
\end{equation}
We have taken the electric field to be polarized along the $x$ direction and using the relations $\bra{\{0\}}\hat{a}^+_{\{k'\}}\ket{\{1\}_{\{k\}}}=0$, and $\bra{\{0\}}\hat{a}_{\{k'\}}\ket{\{1\}_{\{k\}}}=\delta_{\{k'\}\{k\}}$ we obtain
\begin{eqnarray}\label{8}
\mathrm{T}^{\{k\}}_{if}&=&-\frac{1}{2}\sum_{i}\bra{e}\hat{Q}_{ix}\ket{g} \bra{\{0\}} \frac{\partial  \hat{E_x}}{\partial R_i} \ket{\{1\}_{\{k\}}} \nonumber \\
&=&\hbar\Omega^{Q}_{\{k\}}(\mathbf{R})e^{i\theta_{\{k\}}(\mathbf{R})}
\end{eqnarray}
where  $\Omega^{Q}_{\{k\}}(\mathbf{R})$ is the quadrupole Rabi frequency. 
The final state of the system in the absorption process consists of a continuous band of energy of width $\hbar\gamma$ where $\gamma $ is the spontaneous emission rate in free space.  In this case    the absorption rate is given by the form of Fermi's golden rule \cite{Barnett2002,loudon2000quantum,Lloyd2012,fox2006quantum} with a density of states 

\begin{eqnarray}\label{9}
    \mathrm{\Gamma}_{if}&=&\frac{2\pi}{\hbar^2}\big |\mathrm{T}^{\{k\}}_{if}\big |^2\mathcal{F}_{\omega_a}(\omega) \nonumber\\
&=& 2\pi\big |\Omega^{Q}_{\{k\}}(\mathbf{R})\big|^2\mathcal{F}_{\omega_a}(\omega),
\end{eqnarray}
where the density of states is such that  $\mathcal{F}_{\omega_a}(\omega) d\omega$ is the number of upper atomic states that fall within the frequency range $\omega_a$ to $\omega_a + d\omega$. 
The density of final states is a function that peaks at the line centre defined by $\hbar \omega_a= \mathcal{E}_2-\mathcal{E}_1$ and is normalized so that
\begin{equation}\label{9pp}
\int_{-\infty}^\infty\mathcal{F}_{\omega_a}(\omega)d\omega=1.
\end{equation}
The density of states is represented well by a Lorentzian distribution of states with a width (FWHM) coinciding with the spontaneous quadrupole emission rate, thus
\begin{equation}\label{9ppp}
\mathcal{F}_{\omega_a}(\omega)= \frac{1}{\pi}\frac{\gamma/2}{(\omega-\omega_a)^2+(\gamma/2)^2},
\end{equation}
This function representing the density of states  provides a limit to the validity of using Fermi's Golden rule to evaluate the absorption rate, since such a rate is valid only if  the frequency  width of the upper state $\ket{e}$ is larger than the excitation rate; i.e.,the spontaneous emission rate is larger than the Rabi frequency. For high intensities, the Rabi frequency may exceed the spontaneous emission rate, in which case the perturbative approach culminating in the Fermi Golden Rule is no longer valid and the strong coupling regime is applicable involving Rabi oscillations.
The maximum value of the density of states  as a function of $\omega$ is  $\frac {2}{\pi \gamma }$, located at $\omega=\omega_a$.
Substituting Eq. (\ref{9ppp}) in Eq. (\ref{9}) we find for  the quadrupole absorption rate
\begin{equation}\label{main}
\Gamma_{if}=  \frac{\gamma}{(\omega-\omega_a)^2+(\gamma/2)^2}\big |\Omega^{Q}_{\{k\}}(\mathbf{R})\big|^2
\end{equation}
We can now proceed to evaluate the absorption rate when the optical vortex is a Laguerre-Gaussian (LG) mode.   

\section{Absorption of a Laguerre-Gaussian Mode}\label{sec4} 
In the paraxial regime the quadrupole Rabi frequency associated with the LG$_{\ell p}$  of frequency $\omega$, which is plane-polarised along the $x-$direction can be written as follows  \cite{andrews2011structured, babiker2018atoms, klimov1996quadrupole,lin2016dielectric, fickler2012quantum, Domokos2003, deng2008propagation}
\begin{equation}\label{12}
    \Omega _{k\ell p}^{Q} (\rho)=\left(u_{p}^{\ell } (\rho)/\hbar \right)\left (U({\bf R}) Q_{xx} +V({\bf R}) Q_{yx} +ik Q_{zx} \right)
\end{equation}
where the functions $U({\bf R})$ and $V({\bf R})$ are as follows
\begin{eqnarray}
U({\bf R}) =\left(\frac{\left|\ell \right|X}{\rho^{2} } -\frac{2X}{w_{0}^{2} } -\frac{i\ell Y}{\rho^{2} } +\frac{1}{L_{p}^{\left|\ell \right|} } \frac{\partial L_{p}^{\left|\ell \right|} }{\partial X} \right) \label{17a}\label{13},\\
V({\bf R}) =\left(\frac{\left|\ell \right|Y}{\rho^{2} } -\frac{2Y}{w_{0}^{2} } +\frac{i\ell X}{\rho^{2} } +\frac{1}{L_{p}^{\left|\ell \right|} } \frac{\partial L_{p}^{\left|\ell \right|} }{\partial Y} \right) \label{14},
\end{eqnarray}
and
\begin{equation}\label{15}
    u_{\{k\}}(\rho)=u_{k\ell p}(\rho)=E_{k00}f_{\ell,p}(\rho)
\end{equation}
with
\begin{equation}\label{15p}
f_{\ell,p}(\rho)=\sqrt{\frac{p!}{(|\ell|+p)!}}\Big( \frac{\rho\sqrt{2}}{w_0}\Big)^{|\ell|}L_p^{|\ell|}(\frac{2\rho^2}{w_0^2})e^{-\rho^2/w_0^2},
\end{equation}
where $L_p^{|\ell|}$ is the associated Laguerre polynomial  and $w_0$ is the radius at beam waist  (at $Z=0$).  The overall factor $E_{k00}$ is the constant amplitude of the corresponding  plane electromagnetic wave.
The phase function of the LG mode in the paraxial regime is as follows
\begin{equation}\label{16}
\theta_{klp}(\rho,Z,t)\approx kZ+l\phi-\omega t .
\end{equation}

Substituting in Eq.(\ref{main}) we have the quadrupole absorption rate for an atom interacting with the LG$_{\ell,p}$ light mode that is polarized along the x-direction and the atom is characterised by the three quadrupole matrix elements $Q_{xx}, Q_{xy}$ and $Q_{xz}$
\begin{widetext}
\begin{equation}\label{result}
\Gamma_{if}=  \frac{\gamma}{(\omega-\omega_a)^2+(\gamma/2)^2}\left|\left (U({\bf R})Q_{xx} +V({\bf R}) Q_{yx} +ik Q_{zx} \right)\right|^2\times \left|u_{p}^{\ell } (\rho)/\hbar \right|^2.
\end{equation}
\end{widetext}
So far the treatment is general and Eq.(\ref{result}) is the main result of this paper.  This result applies to any atom with a dipole-forbidden but quadrupole-allowed transition which  is at near resonance with a linearly polarized Laguerre-Gaussian light mode LG $_{\ell,p}$.  The main requirement is that the interaction must conform with the OAM selection rules involving the quantum number $m$ between the ground and excited atomic states $\ket{g}$ and $\ket{e}$ and we have for a quadrupole transition
\begin{equation}
\Delta m=0,\pm 1,\pm 2
\end{equation}
The requirement of OAM conservation then means that the optical vortex absorption process in a quadrupole transition can only occur for optical vortices with winding numbers $\ell=0,+1,+2$.
The case $\ell=0$ is possible, but then no transfer of OAM occurs in the absorption process, while each of the cases $\ell=1$ and $\ell=2$ are accompanied by a transfer of OAM of magnitudes $\hbar$ and $2\hbar$, respectively. The details will depend on the specific atom and its specific quadrupole transition. Note that although the radial quantum number $p$ is important for the amplitude distribution function of the LG$_{\ell,p}$
mode, the magnitude of the OAM transferred is determined solely by the value of the winding number $\ell\leq 2$.  

In order to illustrate the main result with practical examples, we may focus on the case that has recently been discussed \cite{Bougouffa20,babiker2018atoms,lembessis2013enhanced,Al-Awfi2019}, namely an LG mode of the winding number $\ell=0,1,2$ and radial number $p$. In the simplest case where the mode is a doughnut mode $p=0$ we find that the last terms involving the derivatives in $U({\bf R})$ and $V({\bf R})$ given by Eqs. (\ref{13},\ref{14}) vanish, as $L_{0}^{\left|\ell \right|}$ are constants for all $\ell$. However the case where $p\neq 0$ is of also of interest since the value of $p$ is important for the intensity distribution.  A specific atomic transition we shall consider to illustrate the results is that of the neutral caesium atom, namely $6^2S_{1/2}\rightarrow 5^2D_{5/2}$ transition.  However, in order to proceed with evaluations, we need the values of the quadrupole matrix elements $Q_{xx}, Q_{xy}$ and $Q_{xz}$ applicable in the transition, depending on the OAM selection rules. 

\section{Vortex absorption in Caesium}\label{sec5}
 The quadrupole matrix elements  $Q_{xx}, Q_{xy}$ and $Q_{xz}$ can be discussed with reference to the normalized hydrogen-like wave function $\psi_{nLm}$ \cite{Bransden2003,Fischer1973} 
 \begin{widetext}
 \begin{equation}\label{29}
\psi_{nLm}(r,\theta, \phi)=\Big\{ \big (\frac{2Z_a}{na_{\mu}}\big)^3 \frac{(n-L-1)!}{2n(n+L)!}\Big\}^{1/2}e^{-\rho(r)/2}\rho(r)^l \mathrm{L}_{n-L-1}^{2L+1}(\rho(r))Y_{L}^{m}(\theta, \phi)
\end{equation}
\end{widetext}
where $a_{\mu}=\frac{4\pi\epsilon_0\hbar^2}{\mu e^2}=a_0\frac{m_e}{\mu}$ is the reduced Bohr radius, $\mathrm{L}_{n-L-1}^{2L+1}(\rho(r))$ are the associated Laguerre polynomials, and $\rho(r)=\frac{2Z_a r}{na_{\mu}}$. The valence electron of Cs atom sees an effective nuclear charge of $Z_a=8.56$ \cite{Bransden2003,Fischer1973,LeKien2018,Varshalovich1988,Yannopapas2015,Ray2020a}. Using Eq.(\ref{29}), we have for the electric quadrupole matrix element  
\begin{equation}\label{29a}
Q_{\alpha \beta} =e\bra {\psi_f} x_{\alpha}x_{\beta} \ket {\psi_i},
\end{equation} 
where $x_{\alpha}=(x,y,z)$. Straightforward evaluations yield the following:
\begin{itemize}
  \item for the case $\Delta m=0$, we find that  $Q_{xx}=\frac{70.2}{Z_a^2}ea_\mu ^2 $
 and $Q_{xy}=Q_{xz}=0$,
  \item  for the case $\Delta m=\pm1 $, we have $Q_{zx}=i\frac{86}{Z_a^2}ea_\mu ^2$ and $Q_{xx}=Q_{yx}=0$,
  \item for $\Delta m=\pm 2$, we have $Q_{xx}=\pm i Q_{yx}=\frac{86}{Z_a^2}ea_\mu ^2$ and $Q_{zx}=0$.
\end{itemize}

\begin{figure}[h]
 \includegraphics[width=0.8\linewidth,height=0.5\linewidth]{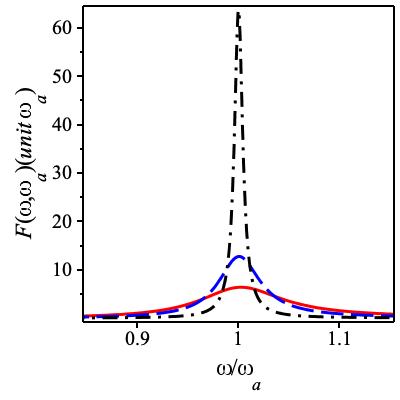}
\caption{(Color online) The normalized modified density function $F(\omega,\omega_a)$[unit $\omega_a$] as a function of the ratio $\omega/\omega_a$. The red solid curve is for $\gamma/\omega_a=0.1$, the blue dashed curve is for  $\gamma/\omega_a=0.05$ and the black dash-dotted curve is for  $\gamma/\omega_a=0.01$. Note the tendency of $F(\omega,\omega_a)$ to become a function proportional to a delta-function as the value of $\gamma/\omega_a$ decreases}\label{FD}
\end{figure}

We consider a quadrupole  transition with the selection rule $\Delta m=1$ applicable for the $(6^2S_{1/2}\rightarrow5^2D_{5/2})$ quadrupole transition in Cs. In this case, The quadrupole moments are  $Q_{xx} =Q_{xy}=0$ and $Q_{xz}\neq 0$ and the Rabi frequency Eq. (\ref{12}) is as follows:
\begin{equation}\label{17p}
    \Omega _{k\ell 0}^{Q} (\rho)=ik Q_{zx}\left (u_{0}^{|\ell| }(\rho)/\hbar \right),
\end{equation}  
The absorption rate is then given by
\begin{eqnarray}\label{20}
\mathrm{\Gamma}_{if}&=&\frac{\gamma}{(\omega-\omega_a)^2+(\gamma/2)^2}\big |\Omega^{Q}_{\{k\}}(\mathbf{R})\big|^2,\\
   &=&\frac{2\pi w_0^2}{c^2}\big | \Omega_0 \big|^{2}F(\omega,\omega_a)\big|f_{\ell,0}(\rho)\big|^2\label{20p}
\end{eqnarray}
where   $f_{\ell,0}$ is given by Eq.(\ref{15p}) with p$=0$  and $\Omega_0$ is a scaling factor for the Rabi frequency
\begin{equation}\label{19p}
\Omega_0=\frac{1}{\hbar}\frac{E_{k00} Q_{zx}}{w_0}
\end{equation}
and the modified density function $F(\omega,\omega_a)$ is given by
\begin{equation}\label{21}
F(\omega,\omega_a)=\frac{\gamma/2}{\pi}\frac{\omega^2}{(\omega-\omega_a)^2+(\gamma /2)^2},
\end{equation}
where the maximum value of this function is $\frac{1}{\pi}(\frac{\omega_a^2}{\gamma/2}+\gamma/2)$,  located at $\omega= \omega_a+\frac{(\gamma/2)^2}{\omega_a}$. It is clear that the dependence of the Rabi frequency on the light frequency affects the transition rate. However, the maximum of the function $F_{\omega_a}(\omega)$ is greater than that of the density of states and its position is shifted away from $\omega_a$. In Figure \ref{FD}, we present the shape of the normalized density of state $F_{\omega_a}(\omega)$ in terms of the ratio $\omega/\omega_a$ for different values of $\gamma/\omega_a$.

Typical parameters in this case are \cite{Chan2016}   $\lambda=685 (nm)$, $Q_{zx}\simeq10e a_0^2$,  and the spontaneous decay rate is $\Gamma_S=3.34\times 10^7 (s^{-1})$ \cite{tojo2004absorption,tojo2005precision}. The beam parameters are chosen such that the beam waist $w_0=\lambda d$, where $d$ is a real number, and the intensity $I=\epsilon_0cE_{k00}^2/2$.  Introducing the dimensionless magnitude of the intensity  ${\cal I}= I/I_0 $, where $I_0=1 Wm^{-2}$, the scaling factor of the Rabi frequency can be written as
\begin{equation}\label{29b}
 \Omega_0=\frac{1}{\hbar}\big(\frac{2I}{\epsilon_0 c}\big)^{1/2}\frac{Q_{zx}}{w_0}=5.14\times 10^{-5}\frac{({\cal I})^{1/2}}{d}\Gamma_S.
\end{equation} 
We must make an appropriate choice of the beam waist and the magnitude of the intensity of the field and ensure that $\Omega_0\ll\Gamma_S$, which is the condition of the validity of the Fermi golden rule.

It is clear that for a weak intensity, we can obtain an absorption rate $\Gamma_{if}$ smaller than the spontaneous transition $\Gamma_S$.  

On the other hand, the Lorentzian density of states is chosen with a width given by the spontaneous emission rate $\gamma=\Gamma_S$, where $\Gamma_S\ll\omega_a$, then the transition rate can be written as 
\begin{equation}
\label{30 }
\Gamma_{if}= 9.64 \times 10^{-23}{\cal I}\left(\frac{\Gamma_S}{2\pi}\right)\frac{\omega^2}{(\omega-\omega_a)^2+(\Gamma_S /2)^2}\big|f_{\ell,0}(\rho)\big|^2.
\end{equation}

\begin{figure}[tbh]
\hspace*{-1.8cm}\textbf{(a)} \hspace*{3.9cm}\textbf{(b)} \
 \includegraphics[width=0.48\linewidth,height=0.45\linewidth]{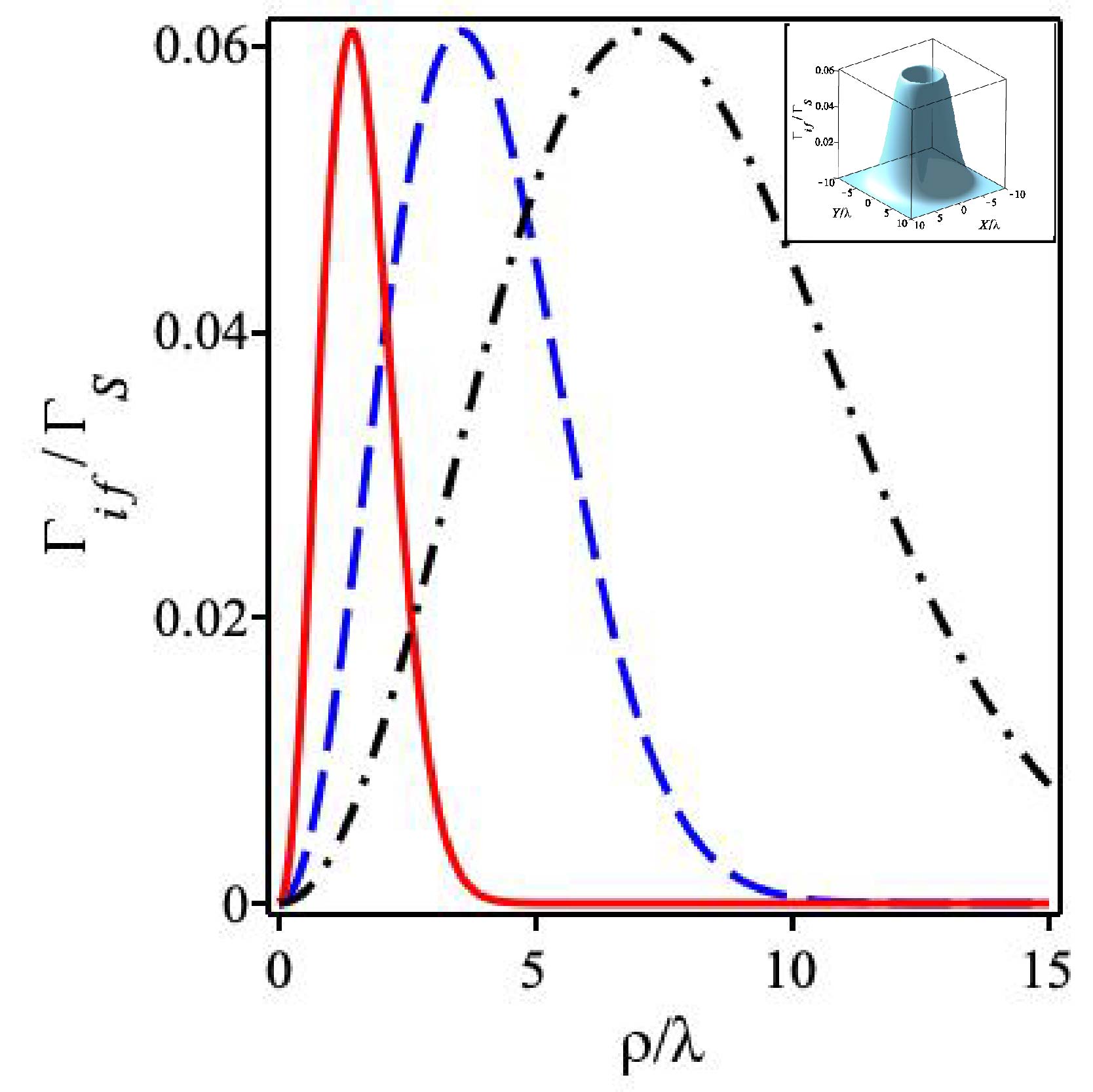}~ \includegraphics[width=0.48\linewidth,height=0.45\linewidth]{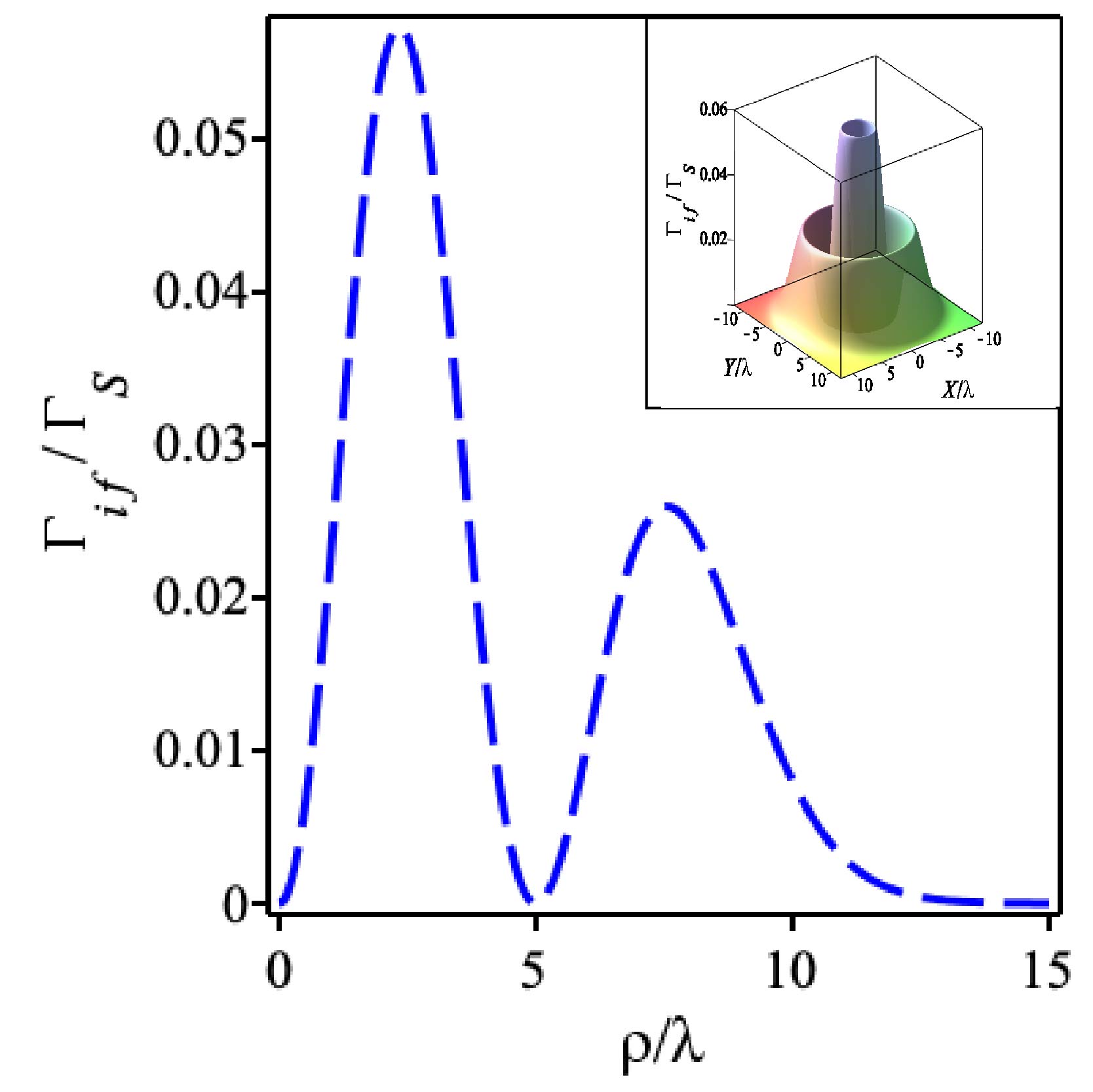}\
\caption{(Color online) The variation with radial position of the quadrupole absorption rate $\Gamma_{if}/\Gamma_S$ for $\Delta m=+1$, (a) $\ell =1, p=0$,  (b) $\ell =1$, $p=1$, and an atom in a Laguerre-Gaussian mode LG$_{\ell,p}$. The red solid line concerns the case $w_0/\lambda=2$; the blue dashed line concerns $w_0/\lambda=5$ and the black dash-dotted line concerns $w_0/\lambda=10$. The insets to the figures represent the cylindrically symmetric quadrupole rate for the case $w_0/\lambda=5$. }\label{Fig1p}
\end{figure}

We assume a moderate laser intensity $I= 40\times 10^{4}Wm^{-2} $ \cite{Chan2016} and substituting for the relevant parameter values we have for the absorption rate at $\omega=\omega_a$
\begin{eqnarray}
\Gamma_{if} & = & 1.67\times 10^{-1}\Gamma_S\big|f_{\ell,0}(\rho)\big|^2, 
\end{eqnarray}
which suggests that the absorption rate is much smaller than the spontaneous rate,  depending on the relative position of the atom.
In Figure \ref{Fig1p}, we present the variation of the absorption rate $\Gamma_{if}/\Gamma_S$ as a function of the radial position of the atom $\rho/\lambda$ for different values of the  beam waist $w_0/\lambda=2, 5, 10$. It is clear that the maximum of the function  shifts away from the origin with increasing beam waist, but the value of the maximum is independent of $w_0$. 

\section{Conclusion}\label{sec6}
This paper has focused on the interaction of atoms with light endowed with orbital angular momentum (OAM) where our main aim was to evaluate the rate of transfer of OAM from the light to the atoms in a dipole-forbidden but quadrupole-allowed transition. Our work follows two significant developments.  The first development concerns the latest experimental confirmation by Giammanco et al \cite{Giammanco2017} following earlier experiments \cite{Araoka2005,loffler2012cholesteric,Loeffler2011a} that OAM cannot be transferred to the internal (electronic-type) degrees of freedom of the atom in an electric dipole transition, though a transfer to the internal degrees of freedom of atoms was predicted theoretically to occur in the normally much weaker quadrupole transition \cite{babiker2018atoms}. The second development concerns the very recent considerable advances in the ability to carry out delicate measurements in experiments targeted specifically at quadrupole transitions involving light carrying OAM \cite{tojo2004absorption, kern2011excitation, cheng2012cavity}.  However, although such experiments have demonstrated the involvement of optical vortices with atoms in quadrupole transitions, the rate of OAM transfer in a quadrupole transition has not, as far as we know, been evaluated, so our task in this work involved setting up the theory of optical vortex photon absorption by an atom in a quadrupole-allowed transition.  Our theory is general and applies to any atom with a quadrupole transition, as for example in Na and Rb atoms, both of which have been the subject of investigations in connection with optical vortex interaction with atoms. However, we proceeded to apply the theory to the well-known case of the Cs $6^2S_{1/2}\rightarrow 5^2D_{5/2}$ quadrupole transition, which conforms with the requirements of OAM conservation consistent with the rules $\Delta m =0,\pm 1, \pm 2$. \\
 Absorption for the case $\Delta m=+1$ required an optical vortex in the form of a Laguerre-Gaussian mode with $\ell=1$ and we have considered two modes, one with $p=0$ and a second with $p=1$.  We have found that the absorption rate as a function of radial position mirrors the intensity distribution of the  beam.  For the doughnut mode the rates peaks at the atomic location of 
$\rho=w_0\sqrt{\ell/2}$, while for the case $p=1$ the maxima are located at $\rho/w_0=\frac{1}{2}\sqrt{2\ell+3 \pm \sqrt{8\ell+9}}$.  Note that the maximum rate is a constant, i.e. independent of $w_0$ and merely shifts its position as $w_0$ increases.\\
The maximum magnitude of the absorption rate in the example we have considered for Cs quadrupole transitions, assuming an intensity of about $10^5 Wm^{-2}$ is of the order of 6\% of the quadrupole spontaneous emission rate.  Such a magnitude should not be beyond the capability of current spectroscopic techniques.

\acknowledgments {The authors are grateful to Professor Stephen Barnett for helpful discussions.}

\bibliographystyle{apsrev4-1}
\bibliography{Ref1}

\begin{thebibliography}{39}%
\makeatletter
\providecommand \@ifxundefined [1]{%
 \@ifx{#1\undefined}
}%
\providecommand \@ifnum [1]{%
 \ifnum #1\expandafter \@firstoftwo
 \else \expandafter \@secondoftwo
 \fi
}%
\providecommand \@ifx [1]{%
 \ifx #1\expandafter \@firstoftwo
 \else \expandafter \@secondoftwo
 \fi
}%
\providecommand \natexlab [1]{#1}%
\providecommand \enquote  [1]{``#1''}%
\providecommand \bibnamefont  [1]{#1}%
\providecommand \bibfnamefont [1]{#1}%
\providecommand \citenamefont [1]{#1}%
\providecommand \href@noop [0]{\@secondoftwo}%
\providecommand \href [0]{\begingroup \@sanitize@url \@href}%
\providecommand \@href[1]{\@@startlink{#1}\@@href}%
\providecommand \@@href[1]{\endgroup#1\@@endlink}%
\providecommand \@sanitize@url [0]{\catcode `\\12\catcode `\$12\catcode
  `\&12\catcode `\#12\catcode `\^12\catcode `\_12\catcode `\%12\relax}%
\providecommand \@@startlink[1]{}%
\providecommand \@@endlink[0]{}%
\providecommand \url  [0]{\begingroup\@sanitize@url \@url }%
\providecommand \@url [1]{\endgroup\@href {#1}{\urlprefix }}%
\providecommand \urlprefix  [0]{URL }%
\providecommand \Eprint [0]{\href }%
\providecommand \doibase [0]{http://dx.doi.org/}%
\providecommand \selectlanguage [0]{\@gobble}%
\providecommand \bibinfo  [0]{\@secondoftwo}%
\providecommand \bibfield  [0]{\@secondoftwo}%
\providecommand \translation [1]{[#1]}%
\providecommand \BibitemOpen [0]{}%
\providecommand \bibitemStop [0]{}%
\providecommand \bibitemNoStop [0]{.\EOS\space}%
\providecommand \EOS [0]{\spacefactor3000\relax}%
\providecommand \BibitemShut  [1]{\csname bibitem#1\endcsname}%
\let\auto@bib@innerbib\@empty
\bibitem [{\citenamefont {Yao}\ and\ \citenamefont {Padgett}(2011)}]{Yao2011a}%
  \BibitemOpen
  \bibfield  {author} {\bibinfo {author} {\bibfnamefont {A.~M.}\ \bibnamefont
  {Yao}}\ and\ \bibinfo {author} {\bibfnamefont {M.~J.}\ \bibnamefont
  {Padgett}},\ }\href@noop {} {\bibfield  {journal} {\bibinfo  {journal} {Adv.
  Opt. Photonics}\ }\textbf {\bibinfo {volume} {3}},\ \bibinfo {pages} {161}
  (\bibinfo {year} {2011})}\BibitemShut {NoStop}%
\bibitem [{\citenamefont {Torres}\ and\ \citenamefont
  {Torner}(2011)}]{TorresTorner2011}%
  \BibitemOpen
  \bibfield  {author} {\bibinfo {author} {\bibfnamefont {J.~P.}\ \bibnamefont
  {Torres}}\ and\ \bibinfo {author} {\bibfnamefont {L.}~\bibnamefont
  {Torner}},\ }\href@noop {} {\emph {\bibinfo {title} {Twisted photons:
  applications of light with orbital angular momentum.}}},\ edited by\ \bibinfo
  {editor} {\bibfnamefont {e.}~\bibnamefont {Torres~JP}, \bibfnamefont
  {Torner~L}}\ (\bibinfo  {publisher} {John Wiley \& Sons.},\ \bibinfo {year}
  {2011})\BibitemShut {NoStop}%
\bibitem [{\citenamefont {Surzhykov}\ \emph {et~al.}(2015)\citenamefont
  {Surzhykov}, \citenamefont {Seipt}, \citenamefont {Serbo},\ and\
  \citenamefont {Fritzsche}}]{Surzhykov2015}%
  \BibitemOpen
  \bibfield  {author} {\bibinfo {author} {\bibfnamefont {A.}~\bibnamefont
  {Surzhykov}}, \bibinfo {author} {\bibfnamefont {D.}~\bibnamefont {Seipt}},
  \bibinfo {author} {\bibfnamefont {V.~G.}\ \bibnamefont {Serbo}}, \ and\
  \bibinfo {author} {\bibfnamefont {S.}~\bibnamefont {Fritzsche}},\ }\href
  {\doibase 10.1103/physreva.91.013403} {\bibfield  {journal} {\bibinfo
  {journal} {Phys. Rev. A}\ }\textbf {\bibinfo {volume} {91}} (\bibinfo {year}
  {2015}),\ 10.1103/physreva.91.013403}\BibitemShut {NoStop}%
\bibitem [{\citenamefont {Metcalf}\ and\ \citenamefont {van~der
  Straten}()}]{metcalflaser}%
  \BibitemOpen
  \bibfield  {author} {\bibinfo {author} {\bibfnamefont {H.}~\bibnamefont
  {Metcalf}}\ and\ \bibinfo {author} {\bibfnamefont {P.}~\bibnamefont {van~der
  Straten}},\ }\href@noop {} {\emph {\bibinfo {title} {Laser Cooling and
  Trapping, Springer Verlag, New York (1999).}}}\BibitemShut {Stop}%
\bibitem [{\citenamefont {Franke-Arnold}\ \emph {et~al.}(2008)\citenamefont
  {Franke-Arnold}, \citenamefont {Allen},\ and\ \citenamefont
  {Padgett}}]{FrankeArnold2008}%
  \BibitemOpen
  \bibfield  {author} {\bibinfo {author} {\bibfnamefont {S.}~\bibnamefont
  {Franke-Arnold}}, \bibinfo {author} {\bibfnamefont {L.}~\bibnamefont
  {Allen}}, \ and\ \bibinfo {author} {\bibfnamefont {M.}~\bibnamefont
  {Padgett}},\ }\href@noop {} {\bibfield  {journal} {\bibinfo  {journal} {Laser
  \& Photonics Reviews}\ }\textbf {\bibinfo {volume} {2}},\ \bibinfo {pages}
  {299} (\bibinfo {year} {2008})}\BibitemShut {NoStop}%
\bibitem [{\citenamefont {Babiker}\ \emph {et~al.}(2019)\citenamefont
  {Babiker}, \citenamefont {Andrews},\ and\ \citenamefont
  {Lembessis}}]{babiker2018atoms}%
  \BibitemOpen
  \bibfield  {author} {\bibinfo {author} {\bibfnamefont {M.}~\bibnamefont
  {Babiker}}, \bibinfo {author} {\bibfnamefont {D.~L.}\ \bibnamefont
  {Andrews}}, \ and\ \bibinfo {author} {\bibfnamefont {V.}~\bibnamefont
  {Lembessis}},\ }\href@noop {} {\bibfield  {journal} {\bibinfo  {journal} {J.
  Opt.}\ }\textbf {\bibinfo {volume} {21}},\ \bibinfo {pages} {013001}
  (\bibinfo {year} {2019})}\BibitemShut {NoStop}%
\bibitem [{\citenamefont {Lembessis}\ \emph {et~al.}(2014)\citenamefont
  {Lembessis}, \citenamefont {Ellinas}, \citenamefont {Babiker},\ and\
  \citenamefont {Al-Dossary}}]{lembessis2014atom}%
  \BibitemOpen
  \bibfield  {author} {\bibinfo {author} {\bibfnamefont {V.}~\bibnamefont
  {Lembessis}}, \bibinfo {author} {\bibfnamefont {D.}~\bibnamefont {Ellinas}},
  \bibinfo {author} {\bibfnamefont {M.}~\bibnamefont {Babiker}}, \ and\
  \bibinfo {author} {\bibfnamefont {O.}~\bibnamefont {Al-Dossary}},\
  }\href@noop {} {\bibfield  {journal} {\bibinfo  {journal} {Phys. Rev. A}\
  }\textbf {\bibinfo {volume} {89}},\ \bibinfo {pages} {053616} (\bibinfo
  {year} {2014})}\BibitemShut {NoStop}%
\bibitem [{\citenamefont {Al-Awfi}\ and\ \citenamefont
  {Bougouffa}(2019)}]{Al-Awfi2019}%
  \BibitemOpen
  \bibfield  {author} {\bibinfo {author} {\bibfnamefont {S.}~\bibnamefont
  {Al-Awfi}}\ and\ \bibinfo {author} {\bibfnamefont {S.}~\bibnamefont
  {Bougouffa}},\ }\href@noop {} {\bibfield  {journal} {\bibinfo  {journal}
  {Results Phys.}\ }\textbf {\bibinfo {volume} {12}},\ \bibinfo {pages} {1357}
  (\bibinfo {year} {2019})}\BibitemShut {NoStop}%
\bibitem [{\citenamefont {Ray}\ \emph {et~al.}(2020)\citenamefont {Ray},
  \citenamefont {Gupta}, \citenamefont {Gokhroo}, \citenamefont {Everett},
  \citenamefont {Nieddu}, \citenamefont {Rajasree},\ and\ \citenamefont
  {Chormaic}}]{Ray2020a}%
  \BibitemOpen
  \bibfield  {author} {\bibinfo {author} {\bibfnamefont {T.}~\bibnamefont
  {Ray}}, \bibinfo {author} {\bibfnamefont {R.~K.}\ \bibnamefont {Gupta}},
  \bibinfo {author} {\bibfnamefont {V.}~\bibnamefont {Gokhroo}}, \bibinfo
  {author} {\bibfnamefont {J.~L.}\ \bibnamefont {Everett}}, \bibinfo {author}
  {\bibfnamefont {T.~N.}\ \bibnamefont {Nieddu}}, \bibinfo {author}
  {\bibfnamefont {K.~S.}\ \bibnamefont {Rajasree}}, \ and\ \bibinfo {author}
  {\bibfnamefont {S.~N.}\ \bibnamefont {Chormaic}},\ }\href {\doibase
  10.1088/1367-2630/ab8265} {\bibfield  {journal} {\bibinfo  {journal} {New J.
  Phys.}\ } (\bibinfo {year} {2020}),\ 10.1088/1367-2630/ab8265}\BibitemShut
  {NoStop}%
\bibitem [{\citenamefont {Bougouffa}\ and\ \citenamefont
  {Babiker}(2020)}]{Bougouffa20}%
  \BibitemOpen
  \bibfield  {author} {\bibinfo {author} {\bibfnamefont {S.}~\bibnamefont
  {Bougouffa}}\ and\ \bibinfo {author} {\bibfnamefont {M.}~\bibnamefont
  {Babiker}},\ }\href {\doibase 10.1103/PhysRevA.101.043403} {\bibfield
  {journal} {\bibinfo  {journal} {Phys. Rev. A}\ }\textbf {\bibinfo {volume}
  {101}},\ \bibinfo {pages} {043403} (\bibinfo {year} {2020})}\BibitemShut
  {NoStop}%
\bibitem [{\citenamefont {van Enk}\ and\ \citenamefont
  {Nienhuis}(1990)}]{van1990light}%
  \BibitemOpen
  \bibfield  {author} {\bibinfo {author} {\bibfnamefont {S.~J.}\ \bibnamefont
  {van Enk}}\ and\ \bibinfo {author} {\bibfnamefont {G.}~\bibnamefont
  {Nienhuis}},\ }\href@noop {} {\bibfield  {journal} {\bibinfo  {journal}
  {Phys. Rev. A}\ }\textbf {\bibinfo {volume} {41}},\ \bibinfo {pages} {3757}
  (\bibinfo {year} {1990})}\BibitemShut {NoStop}%
\bibitem [{\citenamefont {Van~Enk}(1994)}]{van1994selection}%
  \BibitemOpen
  \bibfield  {author} {\bibinfo {author} {\bibfnamefont {S.}~\bibnamefont
  {Van~Enk}},\ }\href@noop {} {\bibfield  {journal} {\bibinfo  {journal}
  {Quantum Optics: Journal of the European Optical Society Part B}\ }\textbf
  {\bibinfo {volume} {6}},\ \bibinfo {pages} {445} (\bibinfo {year}
  {1994})}\BibitemShut {NoStop}%
\bibitem [{\citenamefont {Babiker}\ \emph {et~al.}(2002)\citenamefont
  {Babiker}, \citenamefont {Bennett}, \citenamefont {Andrews},\ and\
  \citenamefont {Romero}}]{29}%
  \BibitemOpen
  \bibfield  {author} {\bibinfo {author} {\bibfnamefont {M.}~\bibnamefont
  {Babiker}}, \bibinfo {author} {\bibfnamefont {C.}~\bibnamefont {Bennett}},
  \bibinfo {author} {\bibfnamefont {D.}~\bibnamefont {Andrews}}, \ and\
  \bibinfo {author} {\bibfnamefont {L.~D.}\ \bibnamefont {Romero}},\
  }\href@noop {} {\bibfield  {journal} {\bibinfo  {journal} {Phys. Rev. Lett.}\
  }\textbf {\bibinfo {volume} {89}},\ \bibinfo {pages} {143601} (\bibinfo
  {year} {2002})}\BibitemShut {NoStop}%
\bibitem [{\citenamefont {Araoka}\ \emph {et~al.}(2005)\citenamefont {Araoka},
  \citenamefont {Verbiest}, \citenamefont {Clays},\ and\ \citenamefont
  {Persoons}}]{Araoka2005}%
  \BibitemOpen
  \bibfield  {author} {\bibinfo {author} {\bibfnamefont {F.}~\bibnamefont
  {Araoka}}, \bibinfo {author} {\bibfnamefont {T.}~\bibnamefont {Verbiest}},
  \bibinfo {author} {\bibfnamefont {K.}~\bibnamefont {Clays}}, \ and\ \bibinfo
  {author} {\bibfnamefont {A.}~\bibnamefont {Persoons}},\ }\href@noop {}
  {\bibfield  {journal} {\bibinfo  {journal} {Phys. Rev. A}\ }\textbf {\bibinfo
  {volume} {71}},\ \bibinfo {pages} {055401} (\bibinfo {year}
  {2005})}\BibitemShut {NoStop}%
\bibitem [{\citenamefont {L{\"o}ffler}\ and\ \citenamefont
  {Woerdman}(2012)}]{loffler2012cholesteric}%
  \BibitemOpen
  \bibfield  {author} {\bibinfo {author} {\bibfnamefont {W.}~\bibnamefont
  {L{\"o}ffler}}\ and\ \bibinfo {author} {\bibfnamefont {J.}~\bibnamefont
  {Woerdman}},\ }in\ \href@noop {} {\emph {\bibinfo {booktitle} {Complex Light
  and Optical Forces VI}}},\ Vol.\ \bibinfo {volume} {8274}\ (\bibinfo
  {organization} {International Society for Optics and Photonics},\ \bibinfo
  {year} {2012})\ p.\ \bibinfo {pages} {827404}\BibitemShut {NoStop}%
\bibitem [{\citenamefont {L{\"o}ffler}\ \emph {et~al.}(2011)\citenamefont
  {L{\"o}ffler}, \citenamefont {Broer},\ and\ \citenamefont
  {Woerdman}}]{Loeffler2011a}%
  \BibitemOpen
  \bibfield  {author} {\bibinfo {author} {\bibfnamefont {W.}~\bibnamefont
  {L{\"o}ffler}}, \bibinfo {author} {\bibfnamefont {D.}~\bibnamefont {Broer}},
  \ and\ \bibinfo {author} {\bibfnamefont {J.}~\bibnamefont {Woerdman}},\
  }\href@noop {} {\bibfield  {journal} {\bibinfo  {journal} {Phys. Rev. A}\
  }\textbf {\bibinfo {volume} {83}},\ \bibinfo {pages} {065801} (\bibinfo
  {year} {2011})}\BibitemShut {NoStop}%
\bibitem [{\citenamefont {Giammanco}\ \emph {et~al.}(2017)\citenamefont
  {Giammanco}, \citenamefont {Perona}, \citenamefont {Marsili}, \citenamefont
  {Conti}, \citenamefont {Fidecaro}, \citenamefont {Gozzini},\ and\
  \citenamefont {Lucchesini}}]{Giammanco2017}%
  \BibitemOpen
  \bibfield  {author} {\bibinfo {author} {\bibfnamefont {F.}~\bibnamefont
  {Giammanco}}, \bibinfo {author} {\bibfnamefont {A.}~\bibnamefont {Perona}},
  \bibinfo {author} {\bibfnamefont {P.}~\bibnamefont {Marsili}}, \bibinfo
  {author} {\bibfnamefont {F.}~\bibnamefont {Conti}}, \bibinfo {author}
  {\bibfnamefont {F.}~\bibnamefont {Fidecaro}}, \bibinfo {author}
  {\bibfnamefont {S.}~\bibnamefont {Gozzini}}, \ and\ \bibinfo {author}
  {\bibfnamefont {A.}~\bibnamefont {Lucchesini}},\ }\href@noop {} {\bibfield
  {journal} {\bibinfo  {journal} {Opt. Lett.}\ }\textbf {\bibinfo {volume}
  {42}},\ \bibinfo {pages} {219} (\bibinfo {year} {2017})}\BibitemShut
  {NoStop}%
\bibitem [{\citenamefont {Lembessis}\ and\ \citenamefont
  {Babiker}(2013)}]{lembessis2013enhanced}%
  \BibitemOpen
  \bibfield  {author} {\bibinfo {author} {\bibfnamefont {V.}~\bibnamefont
  {Lembessis}}\ and\ \bibinfo {author} {\bibfnamefont {M.}~\bibnamefont
  {Babiker}},\ }\href@noop {} {\bibfield  {journal} {\bibinfo  {journal} {Phys.
  Rev. Lett.}\ }\textbf {\bibinfo {volume} {110}},\ \bibinfo {pages} {083002}
  (\bibinfo {year} {2013})}\BibitemShut {NoStop}%
\bibitem [{\citenamefont {Forbes}\ and\ \citenamefont
  {Andrews}(2018)}]{forbes2018chiroptical}%
  \BibitemOpen
  \bibfield  {author} {\bibinfo {author} {\bibfnamefont {K.~A.}\ \bibnamefont
  {Forbes}}\ and\ \bibinfo {author} {\bibfnamefont {D.~L.}\ \bibnamefont
  {Andrews}},\ }in\ \href@noop {} {\emph {\bibinfo {booktitle} {Complex Light
  and Optical Forces XII}}},\ Vol.\ \bibinfo {volume} {10549}\ (\bibinfo
  {organization} {International Society for Optics and Photonics},\ \bibinfo
  {year} {2018})\ p.\ \bibinfo {pages} {1054915}\BibitemShut {NoStop}%
\bibitem [{\citenamefont {Barnett}\ and\ \citenamefont
  {Radmore}(2002)}]{Barnett2002}%
  \BibitemOpen
  \bibfield  {author} {\bibinfo {author} {\bibfnamefont {S.}~\bibnamefont
  {Barnett}}\ and\ \bibinfo {author} {\bibfnamefont {P.~M.}\ \bibnamefont
  {Radmore}},\ }\href@noop {} {\emph {\bibinfo {title} {Methods in theoretical
  quantum optics}}},\ Vol.~\bibinfo {volume} {15}\ (\bibinfo  {publisher}
  {Oxford University Press},\ \bibinfo {year} {2002})\BibitemShut {NoStop}%
\bibitem [{\citenamefont {Loudon}(2000)}]{loudon2000quantum}%
  \BibitemOpen
  \bibfield  {author} {\bibinfo {author} {\bibfnamefont {R.}~\bibnamefont
  {Loudon}},\ }\href@noop {} {\emph {\bibinfo {title} {The quantum theory of
  light}}}\ (\bibinfo  {publisher} {OUP Oxford},\ \bibinfo {year}
  {2000})\BibitemShut {NoStop}%
\bibitem [{\citenamefont {Lloyd}\ \emph {et~al.}(2012)\citenamefont {Lloyd},
  \citenamefont {Babiker},\ and\ \citenamefont {Yuan}}]{Lloyd2012}%
  \BibitemOpen
  \bibfield  {author} {\bibinfo {author} {\bibfnamefont {S.}~\bibnamefont
  {Lloyd}}, \bibinfo {author} {\bibfnamefont {M.}~\bibnamefont {Babiker}}, \
  and\ \bibinfo {author} {\bibfnamefont {J.}~\bibnamefont {Yuan}},\ }\href@noop
  {} {\bibfield  {journal} {\bibinfo  {journal} {Phys. Rev. Lett.}\ }\textbf
  {\bibinfo {volume} {108}},\ \bibinfo {pages} {074802} (\bibinfo {year}
  {2012})}\BibitemShut {NoStop}%
\bibitem [{\citenamefont {Fox}(2006)}]{fox2006quantum}%
  \BibitemOpen
  \bibfield  {author} {\bibinfo {author} {\bibfnamefont {M.}~\bibnamefont
  {Fox}},\ }\href@noop {} {\emph {\bibinfo {title} {Quantum optics: an
  introduction}}},\ Vol.~\bibinfo {volume} {15}\ (\bibinfo  {publisher} {OUP
  Oxford},\ \bibinfo {year} {2006})\BibitemShut {NoStop}%
\bibitem [{\citenamefont {Andrews}(2011)}]{andrews2011structured}%
  \BibitemOpen
  \bibfield  {author} {\bibinfo {author} {\bibfnamefont {D.~L.}\ \bibnamefont
  {Andrews}},\ }\href@noop {} {\emph {\bibinfo {title} {Structured light and
  its applications: An introduction to phase-structured beams and nanoscale
  optical forces}}}\ (\bibinfo  {publisher} {Academic press},\ \bibinfo {year}
  {2011})\BibitemShut {NoStop}%
\bibitem [{\citenamefont {Klimov}\ and\ \citenamefont
  {Letokhov}(1996)}]{klimov1996quadrupole}%
  \BibitemOpen
  \bibfield  {author} {\bibinfo {author} {\bibfnamefont {V.}~\bibnamefont
  {Klimov}}\ and\ \bibinfo {author} {\bibfnamefont {V.}~\bibnamefont
  {Letokhov}},\ }\href@noop {} {\bibfield  {journal} {\bibinfo  {journal}
  {Phys. Rev. A}\ }\textbf {\bibinfo {volume} {54}},\ \bibinfo {pages} {4408}
  (\bibinfo {year} {1996})}\BibitemShut {NoStop}%
\bibitem [{\citenamefont {Lin}\ \emph {et~al.}(2016)\citenamefont {Lin},
  \citenamefont {Jiang}, \citenamefont {Ma}, \citenamefont {Yun}, \citenamefont
  {Liu}, \citenamefont {Werner},\ and\ \citenamefont
  {Mayer}}]{lin2016dielectric}%
  \BibitemOpen
  \bibfield  {author} {\bibinfo {author} {\bibfnamefont {L.}~\bibnamefont
  {Lin}}, \bibinfo {author} {\bibfnamefont {Z.~H.}\ \bibnamefont {Jiang}},
  \bibinfo {author} {\bibfnamefont {D.}~\bibnamefont {Ma}}, \bibinfo {author}
  {\bibfnamefont {S.}~\bibnamefont {Yun}}, \bibinfo {author} {\bibfnamefont
  {Z.}~\bibnamefont {Liu}}, \bibinfo {author} {\bibfnamefont {D.~H.}\
  \bibnamefont {Werner}}, \ and\ \bibinfo {author} {\bibfnamefont {T.~S.}\
  \bibnamefont {Mayer}},\ }\href@noop {} {\bibfield  {journal} {\bibinfo
  {journal} {Appl. Phys. Lett.}\ }\textbf {\bibinfo {volume} {108}},\ \bibinfo
  {pages} {171902} (\bibinfo {year} {2016})}\BibitemShut {NoStop}%
\bibitem [{\citenamefont {Fickler}\ \emph {et~al.}(2012)\citenamefont
  {Fickler}, \citenamefont {Lapkiewicz}, \citenamefont {Plick}, \citenamefont
  {Krenn}, \citenamefont {Schaeff}, \citenamefont {Ramelow},\ and\
  \citenamefont {Zeilinger}}]{fickler2012quantum}%
  \BibitemOpen
  \bibfield  {author} {\bibinfo {author} {\bibfnamefont {R.}~\bibnamefont
  {Fickler}}, \bibinfo {author} {\bibfnamefont {R.}~\bibnamefont {Lapkiewicz}},
  \bibinfo {author} {\bibfnamefont {W.~N.}\ \bibnamefont {Plick}}, \bibinfo
  {author} {\bibfnamefont {M.}~\bibnamefont {Krenn}}, \bibinfo {author}
  {\bibfnamefont {C.}~\bibnamefont {Schaeff}}, \bibinfo {author} {\bibfnamefont
  {S.}~\bibnamefont {Ramelow}}, \ and\ \bibinfo {author} {\bibfnamefont
  {A.}~\bibnamefont {Zeilinger}},\ }\href@noop {} {\bibfield  {journal}
  {\bibinfo  {journal} {Science}\ }\textbf {\bibinfo {volume} {338}},\ \bibinfo
  {pages} {640} (\bibinfo {year} {2012})}\BibitemShut {NoStop}%
\bibitem [{\citenamefont {Domokos}\ and\ \citenamefont
  {Ritsch}(2003)}]{Domokos2003}%
  \BibitemOpen
  \bibfield  {author} {\bibinfo {author} {\bibfnamefont {P.}~\bibnamefont
  {Domokos}}\ and\ \bibinfo {author} {\bibfnamefont {H.}~\bibnamefont
  {Ritsch}},\ }\href@noop {} {\bibfield  {journal} {\bibinfo  {journal} {JOSA
  B}\ }\textbf {\bibinfo {volume} {20}},\ \bibinfo {pages} {1098} (\bibinfo
  {year} {2003})}\BibitemShut {NoStop}%
\bibitem [{\citenamefont {Deng}\ and\ \citenamefont
  {Guo}(2008)}]{deng2008propagation}%
  \BibitemOpen
  \bibfield  {author} {\bibinfo {author} {\bibfnamefont {D.}~\bibnamefont
  {Deng}}\ and\ \bibinfo {author} {\bibfnamefont {Q.}~\bibnamefont {Guo}},\
  }\href@noop {} {\bibfield  {journal} {\bibinfo  {journal} {J. Opt. A: Pure
  Appl. Opt.}\ }\textbf {\bibinfo {volume} {10}},\ \bibinfo {pages} {035101}
  (\bibinfo {year} {2008})}\BibitemShut {NoStop}%
\bibitem [{\citenamefont {Bransden}\ \emph {et~al.}(2003)\citenamefont
  {Bransden}, \citenamefont {Joachain},\ and\ \citenamefont
  {Plivier}}]{Bransden2003}%
  \BibitemOpen
  \bibfield  {author} {\bibinfo {author} {\bibfnamefont {B.~H.}\ \bibnamefont
  {Bransden}}, \bibinfo {author} {\bibfnamefont {C.~J.}\ \bibnamefont
  {Joachain}}, \ and\ \bibinfo {author} {\bibfnamefont {T.~J.}\ \bibnamefont
  {Plivier}},\ }\href@noop {} {\emph {\bibinfo {title} {Physics of atoms and
  molecules}}}\ (\bibinfo  {publisher} {Pearson education},\ \bibinfo {year}
  {2003})\BibitemShut {NoStop}%
\bibitem [{\citenamefont {Fischer}(1973)}]{Fischer1973}%
  \BibitemOpen
  \bibfield  {author} {\bibinfo {author} {\bibfnamefont {C.~F.}\ \bibnamefont
  {Fischer}},\ }\href {\doibase 10.1016/0092-640x(73)90014-4} {\bibfield
  {journal} {\bibinfo  {journal} {Atom. Data Nucl. Data Tabl.}\ }\textbf
  {\bibinfo {volume} {12}},\ \bibinfo {pages} {87} (\bibinfo {year}
  {1973})}\BibitemShut {NoStop}%
\bibitem [{\citenamefont {Le~Kien}\ \emph {et~al.}(2018)\citenamefont
  {Le~Kien}, \citenamefont {Ray}, \citenamefont {Nieddu}, \citenamefont
  {Busch},\ and\ \citenamefont {Chormaic}}]{LeKien2018}%
  \BibitemOpen
  \bibfield  {author} {\bibinfo {author} {\bibfnamefont {F.}~\bibnamefont
  {Le~Kien}}, \bibinfo {author} {\bibfnamefont {T.}~\bibnamefont {Ray}},
  \bibinfo {author} {\bibfnamefont {T.}~\bibnamefont {Nieddu}}, \bibinfo
  {author} {\bibfnamefont {T.}~\bibnamefont {Busch}}, \ and\ \bibinfo {author}
  {\bibfnamefont {S.~N.}\ \bibnamefont {Chormaic}},\ }\href@noop {} {\bibfield
  {journal} {\bibinfo  {journal} {Phys. Rev. A}\ }\textbf {\bibinfo {volume}
  {97}},\ \bibinfo {pages} {013821} (\bibinfo {year} {2018})}\BibitemShut
  {NoStop}%
\bibitem [{\citenamefont {Varshalovich}\ \emph {et~al.}(1988)\citenamefont
  {Varshalovich}, \citenamefont {Moskalev},\ and\ \citenamefont
  {Khersonskii}}]{Varshalovich1988}%
  \BibitemOpen
  \bibfield  {author} {\bibinfo {author} {\bibfnamefont {D.~A.}\ \bibnamefont
  {Varshalovich}}, \bibinfo {author} {\bibfnamefont {A.~N.}\ \bibnamefont
  {Moskalev}}, \ and\ \bibinfo {author} {\bibfnamefont {V.~K.}\ \bibnamefont
  {Khersonskii}},\ }\href {\doibase 10.1142/0270} {\enquote {\bibinfo {title}
  {Quantum theory of angular momentum},}\ } (\bibinfo {year}
  {1988})\BibitemShut {NoStop}%
\bibitem [{\citenamefont {Yannopapas}\ and\ \citenamefont
  {Paspalakis}(2015)}]{Yannopapas2015}%
  \BibitemOpen
  \bibfield  {author} {\bibinfo {author} {\bibfnamefont {V.}~\bibnamefont
  {Yannopapas}}\ and\ \bibinfo {author} {\bibfnamefont {E.}~\bibnamefont
  {Paspalakis}},\ }\href {\doibase 10.1080/09500340.2015.1045435} {\bibfield
  {journal} {\bibinfo  {journal} {J. Mod. Opt.}\ }\textbf {\bibinfo {volume}
  {62}},\ \bibinfo {pages} {1435} (\bibinfo {year} {2015})}\BibitemShut
  {NoStop}%
\bibitem [{\citenamefont {Chan}\ \emph {et~al.}(2016)\citenamefont {Chan},
  \citenamefont {Aljunid}, \citenamefont {Zheludev}, \citenamefont
  {Wilkowski},\ and\ \citenamefont {Ducloy}}]{Chan2016}%
  \BibitemOpen
  \bibfield  {author} {\bibinfo {author} {\bibfnamefont {E.~A.}\ \bibnamefont
  {Chan}}, \bibinfo {author} {\bibfnamefont {S.~A.}\ \bibnamefont {Aljunid}},
  \bibinfo {author} {\bibfnamefont {N.~I.}\ \bibnamefont {Zheludev}}, \bibinfo
  {author} {\bibfnamefont {D.}~\bibnamefont {Wilkowski}}, \ and\ \bibinfo
  {author} {\bibfnamefont {M.}~\bibnamefont {Ducloy}},\ }\href@noop {}
  {\bibfield  {journal} {\bibinfo  {journal} {Opt. Lett.}\ }\textbf {\bibinfo
  {volume} {41}},\ \bibinfo {pages} {2005} (\bibinfo {year}
  {2016})}\BibitemShut {NoStop}%
\bibitem [{\citenamefont {Tojo}\ \emph {et~al.}(2004)\citenamefont {Tojo},
  \citenamefont {Hasuo},\ and\ \citenamefont {Fujimoto}}]{tojo2004absorption}%
  \BibitemOpen
  \bibfield  {author} {\bibinfo {author} {\bibfnamefont {S.}~\bibnamefont
  {Tojo}}, \bibinfo {author} {\bibfnamefont {M.}~\bibnamefont {Hasuo}}, \ and\
  \bibinfo {author} {\bibfnamefont {T.}~\bibnamefont {Fujimoto}},\ }\href@noop
  {} {\bibfield  {journal} {\bibinfo  {journal} {Phys. Rev. Lett.}\ }\textbf
  {\bibinfo {volume} {92}},\ \bibinfo {pages} {053001} (\bibinfo {year}
  {2004})}\BibitemShut {NoStop}%
\bibitem [{\citenamefont {Tojo}\ \emph {et~al.}(2005)\citenamefont {Tojo},
  \citenamefont {Fujimoto},\ and\ \citenamefont {Hasuo}}]{tojo2005precision}%
  \BibitemOpen
  \bibfield  {author} {\bibinfo {author} {\bibfnamefont {S.}~\bibnamefont
  {Tojo}}, \bibinfo {author} {\bibfnamefont {T.}~\bibnamefont {Fujimoto}}, \
  and\ \bibinfo {author} {\bibfnamefont {M.}~\bibnamefont {Hasuo}},\
  }\href@noop {} {\bibfield  {journal} {\bibinfo  {journal} {Phys. Rev. A}\
  }\textbf {\bibinfo {volume} {71}},\ \bibinfo {pages} {012507} (\bibinfo
  {year} {2005})}\BibitemShut {NoStop}%
\bibitem [{\citenamefont {Kern}\ and\ \citenamefont
  {Martin}(2011)}]{kern2011excitation}%
  \BibitemOpen
  \bibfield  {author} {\bibinfo {author} {\bibfnamefont {A.~M.}\ \bibnamefont
  {Kern}}\ and\ \bibinfo {author} {\bibfnamefont {O.~J.}\ \bibnamefont
  {Martin}},\ }\href@noop {} {\bibfield  {journal} {\bibinfo  {journal} {Nano
  Lett.}\ }\textbf {\bibinfo {volume} {11}},\ \bibinfo {pages} {482} (\bibinfo
  {year} {2011})}\BibitemShut {NoStop}%
\bibitem [{\citenamefont {Cheng}\ \emph {et~al.}(2012)\citenamefont {Cheng},
  \citenamefont {Sun}, \citenamefont {Pan}, \citenamefont {Lu}, \citenamefont
  {Li}, \citenamefont {Wang}, \citenamefont {Liu},\ and\ \citenamefont
  {Hu}}]{cheng2012cavity}%
  \BibitemOpen
  \bibfield  {author} {\bibinfo {author} {\bibfnamefont {C.-F.}\ \bibnamefont
  {Cheng}}, \bibinfo {author} {\bibfnamefont {Y.}~\bibnamefont {Sun}}, \bibinfo
  {author} {\bibfnamefont {H.}~\bibnamefont {Pan}}, \bibinfo {author}
  {\bibfnamefont {Y.}~\bibnamefont {Lu}}, \bibinfo {author} {\bibfnamefont
  {X.-F.}\ \bibnamefont {Li}}, \bibinfo {author} {\bibfnamefont
  {J.}~\bibnamefont {Wang}}, \bibinfo {author} {\bibfnamefont {A.-W.}\
  \bibnamefont {Liu}}, \ and\ \bibinfo {author} {\bibfnamefont {S.-M.}\
  \bibnamefont {Hu}},\ }\href@noop {} {\bibfield  {journal} {\bibinfo
  {journal} {Opt. Express}\ }\textbf {\bibinfo {volume} {20}},\ \bibinfo
  {pages} {9956} (\bibinfo {year} {2012})}\BibitemShut {NoStop}%
\end{thebibliography}%

\end{document}